# Phase Field Modeling in Social Media Dynamics: Simulation of Opinion Evolution with Feedback, Separation

Yasuko Kawahata [†]

Faculty of Sociology, Department of Media Sociology, Rikkyo University, 3-34-1 Nishi-Ikebukuro,Toshima-ku, Tokyo, 171-8501, JAPAN.
`ykawahata@rikkyo.ac.jp,kawahata.lab3@damp.tottori-u.ac.jp`

**Abstract:** This study proposes a novel numerical simulation model that represents the degree of understanding and cognition of information on social networks as continuous phase field variables. Information opinions are defined as phase field variables $\phi_A, \phi_B, \phi_C$, modeling the inclination of individual users' opinions. The simulation reflects the characteristics of communication media with immediacy and bidirectionality, specifically social networking services (SNS), dynamically reproducing the propagation of information and feedback mechanisms.The model sets internal judgment conditions as parameters, simulating psychosocial processes such as confirmation bias, social influence, forgetfulness, and opinion rigidity. This allows for a numerical analysis of how individual users process information and how opinions evolve as a result. Furthermore, the model describes the phase separation dynamics of information between filter bubbles and non-bubble regions, detailing the interactions and evolution of opinions at the boundaries of spaces with different information concentrations. The spatial distribution of opinions and their dynamics under conditions where different opinions coexist and interact are simulated from the perspective of phase separation and interaction energy. This research utilizes a phase field model to elucidate the complexities of opinion formation on real-time, bidirectional media like SNS. It quantitatively demonstrates how information spreads and opinions solidify, revealing the mechanisms of opinion evolution inside and outside filter bubbles. By combining theoretical frameworks with observational data from actual social networks, it analyzes the impact of information concentration on opinion evolution and the outcomes of social interactions on opinion distribution. The model aims to provide a foundation for deepening our understanding of significant social phenomena in contemporary digital communication, such as opinion polarization and echo chamber formation on SNS.

## 1. Introduction

The intersection of sociophysics and digital communication has opened a novel vista for understanding human behavior and opinion dynamics. The digital age has transformed how opinions are formed, evolved, and propagated, especially on platforms such as social networking services (SNS). Capturing this complexity requires innovative models that reflect both the immediacy of information exchange and the myriad cognitive processes at play within individuals. This study introduces a comprehensive numerical simulation model that encapsulates these aspects using a phase field approach to represent the understanding and cognition of information as continuous variables.

Our model delineates information opinions as phase field variables $\phi_A, \phi_B, \phi_C$, corresponding to the inclination towards opinions A, B, and C, respectively. These variables enable a dynamic simulation of opinion evolution on SNS by incorporating the immediacy and bidirectionality inherent in digital communication. This approach allows for the real-time propagation of information and its feedback mechanisms, akin to the physical phenomena of phase separation and pattern formation observed in materials science.

Incorporating principles from sociophysics, we draw upon the foundational work of Galem (1982) and subsequent research that utilized statistical physics-based methods to model opinion dynamics. These studies highlighted the existence of critical thresholds in societal behaviors, akin to phase transitions in physical systems. By adapting these concepts, we simulate the internal judgment conditions that govern how individuals process information. These conditions include psychosocial processes such as confirmation bias, social influence, forgetfulness, and opinion rigidity. The model sets these as parameters, offering a numerical analysis of individual and collective opinion evolution.

Our simulation also delves into the dynamics of phase separation, particularly in the context of filter bubbles—a phenomenon prevalent in modern SNS usage. By considering the interactions and evolution of opinions at the boundary between high and low information concentration spaces, we simulate the spatial distribution of opinions and their interplay. This is done through the lens of phase separation and interaction energy, which allows for a granular examination of the conditions under which different opinions coexist and how they influence one another.

The study's foundation rests on the Cahn-Hilliard equation(equation used to describe the dynamics of phase separation and to model the time variation of the concentration fields of the components. It is often used in conjunction with the Allen-Kahn equation, and both are key components of the phase-field model to describe the evolution of multiphase systems.). And Ginzburg-Landau, Allen-Cahn equation, Phase Field Models(This method can be extended to handle phase changes involving order/disorder transitions, and Landau theory is used to describe phase transitions and expresses the free energy as a function of the constitutive variables. It provides the theoretical framework underlying the phase-field model and can handle spatial variations in order parameters. Traditionally used to describe phase interface motion. Also Potts model or Q-state Potts model: a statistical physical model used specifically to simulate the microstructure of alloys and the dynamics of phase boundaries. By extending this equation to accommodate the multi-component system of opinion dynamics, we offer a sophisticated model that captures the nuances of opinion formation and evolution in non-equilibrium social systems. This model takes into account the complexities of individual interactions, providing insight into the mechanisms of opinion polarization and echo chamber formation on SNS.

The significance of this research lies in its potential to deepen our understanding of contemporary social phenomena. By integrating theoretical frameworks with empirical data from SNS, we aim to elucidate the role of information concentration on opinion dynamics and the effects of social interactions on opinion distribution. This endeavor is particularly pertinent in light of recent global events that have highlighted the importance of opinion dynamics in societal transformations and conflicts.

In summary, our study presents a novel approach to modeling opinion dynamics in digital communication, offering a theoretical and computational framework for analyzing the complex interplay of individual cognitive processes and collective behaviors. Through this work, we strive to provide insights into the digital citizenry's opinion formation and evolution, contributing to the broader field of sociophysics and its applications in understanding the digital society.

## 2. Opinion Distribution Preview

Opinion dynamics theory is applied to compute simulations of human behavior in society. Preview our research, we introduce distrust into the bounded trust model in order to discuss the time transition and trust between the two. For a fixed agent, $1 \leq i \leq N$, the agent's opinion at time $t$ is $I_i(t)$. As a trust coefficient, we modified the meaning of the coefficient $D_{ij}$ in the bounded trust model. Here, we assumed that $D_{ij} > 0$ if there is trust between them and $D_{ij} < 0$ if there is distrust between them. For the calculations in this paper, $D_{ij}$ was assumed to be constant. Thus, the change in the opinion of agent $i$ can be expressed as follows:

$$\Delta I_i(t) = -\alpha I_i(t) + c_i A(t)\Delta t + \sum_{j=1}^{N} D_{ij} I_j(t)\Delta t \qquad (1)$$

$$D_{ij}\phi(I_i, I_j)(I_j(t) - I_i(t)) \qquad (2)$$

$$\phi(I_i, I_j) = \frac{1}{1 + \exp(\beta(|I_i - I_j| - b))} \qquad (3)$$

Here, in order to cut off the influence from people whose opinions differ significantly(3), we use the following sigmoidal smooth cutoff function, which is a Fermi function system. In other words, the model hypothesizes that people do not pay attention to opinions that are far from their own. We will have a separate discussion on the introduction regarding this Fermi function system in the future. Here, $D_{ij}$ and $D_{ji}$ are assumed to be independent. Usually, $D_{ij}$ is an asymmetric matrix.

### 2.1 Phase field modeling of information states

The simulation evolves the opinions (or states) of three different populations, denoted as $\phi_A$, $\phi_B$, and $\phi_C$. These opinions are distributed across a spatial grid and are updated over time based on a set of dynamical rules that take into account diffusion, bias, forgetfulness, and the rigidity of opinions.

### Parameters

The following parameters are used in the simulation:

Number of simulation steps: $num\_steps = 100$

Size of spatial grid: $N = 100$

Strength of bias towards a particular opinion: $bias\_strength = 0.1$

Strength of social influence: $social\_influence\_strength = 0.5$

Rate of forgetfulness: $forgetfulness\_strength = 0.1$

Time step width: $\Delta t = 0.01$

Spatial grid width: $\Delta x = 1.0$

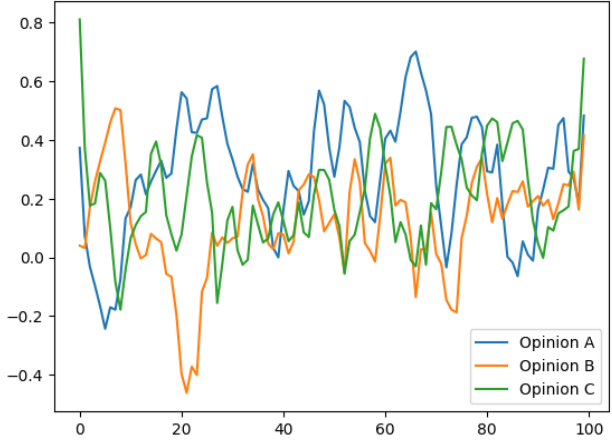

Fig. 1: $\phi_A, \phi_B, \phi_C$, Opinion$_R$igidity, Bias, $t = 100$

## Dynamical Rules

The update rules for the opinions $\phi_A$, $\phi_B$, and $\phi_C$ at each grid point are given by:

$$\frac{\partial \phi_A}{\partial t} = \frac{\phi_A(i-1) - 2\phi_A(i) + \phi_A(i+1)}{\Delta x^2}$$
$$- opinion\_rigidity(i) \cdot (\phi_B(i) + \phi_C(i))$$
$$+ bias(i) - forgetfulness(i) \cdot \phi_A(i)$$
$$\frac{\partial \phi_B}{\partial t} = \frac{\phi_B(i-1) - 2\phi_B(i) + \phi_B(i+1)}{\Delta x^2}$$
$$- opinion\_rigidity(i) \cdot (\phi_A(i) + \phi_C(i))$$
$$- bias(i) - forgetfulness(i) \cdot \phi_B(i)$$
$$\frac{\partial \phi_C}{\partial t} = \frac{\phi_C(i-1) - 2\phi_C(i) + \phi_C(i+1)}{\Delta x^2}$$
$$- opinion\_rigidity(i) \cdot (\phi_A(i) + \phi_B(i))$$
$$- forgetfulness(i) \cdot \phi_C(i)$$

where $i$ ranges from 1 to $N-1$ and represents the spatial grid point. The terms involving $\Delta x^2$ represent diffusion, and $opinion\_rigidity(i)$ represents the rigidity of opinions at each grid point.

## Initial Conditions

Initial states for $\phi_A$, $\phi_B$, and $\phi_C$ are random distributions over the interval $[0, 1]$. A random seed is set to ensure reproducibility of the simulation.

Phase field modeling of information states:

Variables representing information opinions are

$$\phi_A, \phi_B, \phi_C, \ldots \tag{4}$$

Each individual has these variables and their values represent the degree of inclination toward the opinions $\phi_A, \phi_B, \phi_C,$. Information diffusion is represented by the time variation of these phase-field variables.

### Information Propagation and Feedback

Suppose that the information $\phi_A$, $\phi_B$, and $\phi_C$ have different diffusion coefficients. Let $\phi_A > \phi_B > \phi_C$, whereby $\phi_A > \phi_B > \phi_C$ are more diffuse in that order.

Feedback is the mechanism by which one's state changes under the influence of information from others.

**Internal judgment conditions**

Confirmation bias: individuals selectively accept information that is consistent with their own opinions.

Social influence: sets parameters for susceptibility to the opinions of others.

Forgetting: models that information disappears from the individual over time.

Opinion rigidity: sets the conditions that determine whether one adheres to opinion $\phi_A, \phi_B$, or $\phi_C$.

**Modeling phase separation dynamics**

Model information exchange between filter bubbles and non-bubble regions as a phase interface. Considering interaction energy, it simulates the dynamics when different opinions come into contact.

**Dynamics of opinion change**

Computes the spatial distribution of populations with different opinions. Introduces parameters that control the "mixing" of opinions.

When applying the phase-field model to social diffusion of information, the degree of inclination toward the opinions ($\phi_A$, $\phi_B$, $\phi_C$) held by individuals is modeled as the phase-field variables $\phi_A$, $\phi_B$ and $\phi_C$. These variables represent the state of opinion across time and space for each individual. By including internal judgment conditions such as confirmation bias, social influence, forgetting, and opinion rigidity, the dynamics of information diffusion and individual opinion change can be captured.

Phase separation dynamics models the phenomenon of matter separating into different phases (e.g., liquid and gas). Applying this concept to the modeling of opinions, one can simulate "filter bubbles" or phase boundaries of opinions that form between groups with differing opinions.

Interaction energy is used to show the dynamics of tension and competition when individuals with different opinions come into contact. It corresponds to the interfacial energy between the different phases; the higher this energy, the less likely the different opinions are to mix and the more likely bubble formation will be promoted.

**Allen-Kahn Equation** The Allen-Kahn equation is commonly used to model the dynamics of phase separation. This equation is used to describe the evolution of phase boundaries in phase field modeling.

Opinion mixing parameter: This controls how easily opinions are exchanged between individuals. The higher it is,

the more easily opinions are exchanged and the blurrier the phase boundaries.

We simulate the evolution of opinions across a spatial grid. Each point on the grid represents an individual's inclination towards one of three opinions, denoted as $\phi_A$, $\phi_B$, and $\phi_C$.

**Model Parameters**

$N$: Number of spatial grid points.

num_steps: Total number of simulation steps.

bias_strength: Strength of the opinion bias.

social_influence_strength: Strength of social influence on opinion change.

forgetfulness_strength: Rate of forgetting an opinion.

$dt$: Time step width.

$dx$: Spatial grid width.

**Initial Conditions**

The initial state of opinions $\phi_A$, $\phi_B$, and $\phi_C$ is set randomly for each grid point.

**Dynamics Update**

The opinion dynamics are updated according to the following equation for each opinion $\phi$ at every grid point $i$:

$$\frac{d\phi}{dt} = D\frac{d^2\phi}{dx^2} - I(\phi) + B - F\phi \qquad (5)$$

where

$D$ represents the diffusion term, modeling the spread of opinions.

$I(\phi)$ is the interaction energy with other opinions at the interface, computed as opinion_rigidity $\times (\phi_{\text{other}})$.

$B$ represents the bias towards an opinion.

$F$ is the forgetfulness rate.

The simulation iterates over the specified number of steps, updating the opinions according to the dynamics described above.

## 2.2 Opinion Dynamics Update Function

The opinion dynamics update function calculates the new state of opinions in a spatially distributed population. This function takes into account the interaction energy between different opinions and the tendency of opinions to mix.

**Parameters**

$\phi_A, \phi_B, \phi_C$: Arrays representing the distribution of opinions A, B, and C, respectively, over a spatial grid.

interaction_energy: Array representing the interaction energy at the interface between different opinions.

mix_parameter: Array representing the degree to which opinions mix with each other.

$dt$: Scalar representing the time step.

$dx$: Scalar representing the spatial step.

**Dynamics**

The function updates the opinion states according to the following equations:

Mixing effect:
$M_{A,i} = \text{mix\_parameter}[i] \cdot (\phi_{B,i} + \phi_{C,i})$
$M_{B,i} = \text{mix\_parameter}[i] \cdot (\phi_{A,i} + \phi_{C,i})$
$M_{C,i} = \text{mix\_parameter}[i] \cdot (\phi_{A,i} + \phi_{B,i})$

Diffusion (with $Allen - Cahn term$)
$\Delta\phi_{A,i} = \frac{\phi_{A,i-1} - 2\phi_{A,i} + \phi_{A,i+1}}{dx^2} - E_{A,i} + M_{A,i}$
$\Delta\phi_{B,i} = \frac{\phi_{B,i-1} - 2\phi_{B,i} + phi_{B,i+1}}{dx^2} - E_{B,i} + M_{B,i}$
$\Delta\phi_{C,i} = \frac{\phi_{C,i-1} - 2\phi_{C,i} + phi_{C,i+1}}{dx^2} - E_{C,i} + M_{C,i}$

Temporal evolution:
$\phi'_{A,i} = \phi_{A,i} + dt \cdot \Delta\phi_{A,i}$
$\phi'_{B,i} = \phi_{B,i} + dt \cdot \Delta\phi_{B,i}$
$\phi'_{C,i} = \phi_{C,i} + dt \cdot \Delta\phi_{C,i}$

**Return**

The function returns the updated arrays $\phi'_A, \phi'_B, \phi'_C$ after applying the opinion dynamics for a single time step.

**Model Description**

The `update_phi` function simulates the opinion dynamics over a one-dimensional lattice where each site has an opinion value for three distinct choices: A, B, and C. The update rules incorporate several mechanisms: selective acceptance based on confirmation bias, opinion change due to social influence, information loss through forgetfulness, and opinion rigidity.

**Equations and Parameters**

The state of each opinion at site $i$ is updated at each time step based on the following discrete-time dynamical equations:

Social Influence:
$$S_{A,i} = \text{social\_influence}[i] \cdot (\phi_{A,i-1} + \phi_{A,i+1})$$
$$S_{B,i} = \text{social\_influence}[i] \cdot (\phi_{B,i-1} + \phi_{B,i+1})$$
$$S_{C,i} = \text{social\_influence}[i] \cdot (\phi_{C,i-1} + \phi_{C,i+1})$$

Forgetfulness:
$$F_{A,i} = \text{forgetfulness}[i] \cdot \phi_{A,i}$$
$$F_{B,i} = \text{forgetfulness}[i] \cdot \phi_{B,i}$$
$$F_{C,i} = \text{forgetfulness}[i] \cdot \phi_{C,i}$$

Opinion Rigidity:
$$R_{A,i} = \text{opinion\_rigidity}[i] \cdot \phi_{A,i}$$
$$R_{B,i} = \text{opinion\_rigidity}[i] \cdot \phi_{B,i}$$
$$R_{C,i} = \text{opinion\_rigidity}[i] \cdot \phi_{C,i}$$

Diffusion:
$$D_{A,i} = \frac{\phi_{A,i-1} - 2\phi_{A,i} + \phi_{A,i+1}}{dx^2}$$
$$D_{B,i} = \frac{\phi_{B,i-1} - 2\phi_{B,i} + \phi_{B,i+1}}{dx^2}$$
$$D_{C,i} = \frac{\phi_{C,i-1} - 2\phi_{C,i} + \phi_{C,i+1}}{dx^2}$$

Temporal Evolution:
$$\phi_{A,i}^{(new)} = \phi_{A,i} + dt \cdot (D_{A,i} + B_{A,i} + S_{A,i} - F_{A,i} + R_{A,i})$$
$$\phi_{B,i}^{(new)} = \phi_{B,i} + dt \cdot (D_{B,i} + B_{B,i} + S_{B,i} - F_{B,i} + R_{B,i})$$
$$\phi_{C,i}^{(new)} = \phi_{C,i} + dt \cdot (D_{C,i} + B_{C,i} + S_{C,i} - F_{C,i} + R_{C,i})$$

**Parameters**

**bias**: An array representing the strength of confirmation bias for each site.

**social_influence**: An array representing the strength of social influence at each site.

**forgetfulness**: An array indicating the rate at which information is forgotten at each site.

**opinion_rigidity**: An array indicating the degree to which an opinion at a site resists change.

**dt**: The time step for the simulation.

**dx**: The spatial step representing the distance between sites

The `update_opinion_dynamics` function iteratively updates the state of opinions within a population considering various psychological and social factors.

**Parameters and Variables**

- $N$: The number of agents or nodes in the system.
- $dt$: The time step for the simulation.
- $dx$: The spatial resolution for the simulation.
- `threshold`: The threshold for opinion rigidity.
- `bubble_strength`: The strength of the filter bubble effect.
- `bubble_tolerance`: The tolerance within which the filter bubble effect applies.

**Update Equations**

The dynamics of the opinion states are updated according to the following equations:

for $i = 1$ to $N - 1$:

Interface Energy: $E_i = \text{rig}[i](\phi_{B,i} + \phi_{C,i})$

Social Influence: $S_i = \text{soc\_inf}[i](\phi_{B,i} + \phi_{C,i} - 2\phi_{A,i})$

Confirmation Bias: $B_i = \text{bias}[i]\phi_{A,i}$

Opinion Rigidity: $R_i = \text{rig}[i]\phi_{A,i}$ if $\phi_{A,i} > \text{thresh}$ else 0

Filter Bubble: $F_i = (\phi_{A,i-1} + \phi_{A,i+1})\text{bub\_str}$
if $|\phi_{A,i-1} - \phi_{A,i+1}| < \text{bub\_tol}$ else 0

Update Rule: $\phi'_{A,i} = \phi_{A,i} + dt \cdot (E_i + S_i$
$+ B_i + R_i + F_i) \cdot (1 - \text{forget}[i])$

Where the abbreviations are as follows:

- rigid: opinion rigidity
- soc_inf: social influence
- thresh: threshold for opinion rigidity
- bub_str: bubble strength
- bub_tol: bubble tolerance
- forget: forgetfulness

The same set of equations applies to $\phi_B$ and $\phi_C$, with appropriate substitutions.

The function returns the updated opinion arrays and the change in opinions due to each factor.

**Gradient Calculation and Evolution Recording**

- The gradients of the opinion distributions $\phi_A$, $\phi_B$, and $\phi_C$ with respect to the spatial grid are computed at each simulation step. This is achieved by using the `np.gradient` function which approximates the derivative using central differences in the interior and first differences at the boundaries.
- The resulting gradient arrays are stored in corresponding evolution matrices (`phi_A_grad_evolution`, `phi_B_grad_evolution`, `phi_C_grad_evolution`), with each row corresponding to a time step in the simulation.

- The gradient at each step and position represents the rate of change of the opinion value, which can be indicative of the dynamics of opinion formation and spread.
- Mathematically, the gradient at each grid point $i$ and time step $t$ is given by:

$$\text{Gradient at position } i \text{ and time } t: \quad G_{i,t} = \frac{\partial \phi(i,t)}{\partial x}$$

  where $\partial \phi(i,t)/\partial x$ denotes the partial derivative of the opinion value with respect to space at position $i$ and time $t$.
- The absolute value of the gradient is taken to consider the magnitude of change without regard to the direction of change (i.e., whether the opinion is increasing or decreasing at that point).
- The absolute gradients are recorded for visualization purposes. A heatmap can be used to display how the gradients evolve over time, providing insight into the areas with the most dynamic opinion changes.

- After the gradients are calculated for each opinion type ($A$, $B$, and $C$), a heatmap is created for each to visualize the spatial distribution of opinion change intensities over the course of the simulation.
- The heatmaps show the magnitude of the opinion gradients across the spatial domain at each time step, with warmer colors typically representing higher rates of change.

Opinion dynamics over a series of discrete time steps. Below is a detailed explanation of the processes, associated mathematical formulations, and parameters involved. The opinion states $\phi_A$, $\phi_B$, and $\phi_C$ for three different opinions are updated using two distinct functions:

(1) `update_phi` function updates the opinion states based on bias, social influence, forgetfulness, and opinion rigidity.
(2) `update_opinion_dynamics` function then further updates these states considering interaction energies and mixing parameters.

The update rules for the opinion states within each time step $t$ are described by the following equations:

$$\phi'_A, \phi'_B, \phi'_C = \texttt{update\_phi}(\phi_A, \phi_B, \phi_C, \ldots)$$
$$\phi'_A, \phi'_B, \phi'_C = \texttt{update\_op\_dynamics}(\phi'_A, \phi'_B, \phi'_C, \ldots)$$

Each update considers the effects of neighboring opinions and the tendency to maintain one's current state due to rigidity or bias.

Interfaces are calculated to determine the points where opinion changes are significant:

$$\texttt{interfaces}_X = \texttt{calculate\_interface}(\phi_X, \text{threshold})$$

where $X$ can be $A$, $B$, or $C$, representing the different opinions, and `threshold` is a predetermined value that defines what constitutes a significant change in opinion.

After updating the opinions and calculating the interfaces at each step, the simulation results are visualized:

**Process:**

- For each simulation step, the function `Result_histograms` is called to draw histograms of the opinion states ($\phi_A$, $\phi_B$, and $\phi_C$).
- The opinions at each step are added to their respective lists for further analysis.
- After accumulating the data from multiple steps, histograms of all opinion states are plotted to visualize the overall distribution.
- `phi_A_data`, `phi_B_data`, `phi_C_data`: Lists containing the opinion states for each simulation step.
- `bins`: The number of bins or intervals used to divide the range of opinion states for the histogram.

**Equations:**

- Result represent the frequency distribution of opinion states, divided into bins.
- $\text{Frequency}(\phi) = \sum_{\text{all steps}} \text{Count}(\phi \in \text{bin})$

**Cumulative Distribution Function Calculation**

**Process:**

- A cumulative distribution function ($CDF$) is calculated for each opinion state to show the probability that a variable takes a value less than or equal to a certain value.
- The $CDF$ is obtained by normalizing the cumulative sum of the histogram counts.

**Parameters:**

- `all_phi_A`, `all_phi_B`, `all_phi_C`: Lists containing all opinion states across the simulation steps.

- `bins`: Defines the intervals for the histogram, which is used to calculate the $CDF$.

**Equations**

- $CDF(\phi) = \frac{\text{Cumulative Sum}(\text{Count}(\phi \leq x))}{\text{Total Count}}$
- The $CDF$ is as a function of opinion states.
- Result(bin edges, $CDF$)

**Parameters**

The parameters used in the simulation include:

- `bias` - Represents the confirmation bias for each opinion.
- `social_influence` - Measures the social influence affecting each opinion.
- `forgetfulness` - Accounts for the tendency to forget or abandon an opinion.
- `opinion_rigidity` - Reflects the resistance to change one's opinion.
- `interaction_energy` - Quantifies the energetic cost or benefit of having adjacent differing opinions.
- `mix_parameter` - Controls the degree to which opinions mix or influence each other.
- `dt` - The time step size.
- `dx` - The spatial resolution of the simulation.

These parameters are adjusted to model various scenarios of opinion dynamics within a population.

As a final step, the CDF of each process is derived. This experiment function `update_opinion_dynamics` simulates the evolution of opinions within a population. It considers the effects of interaction energy, social influence, confirmation bias, opinion rigidity, and filter bubbles.

- $\phi_A, \phi_B, \phi_C$: Current opinion states for groups A, B, and C.
- `opinion_rigidity`: Resistance to change in opinion.
- `bias`: Tendency to favor an opinion.
- `social_influence`: Effect of society on individual opinion.
- `forgetfulness`: Likelihood of changing opinion over time.
- `dt`: Time step for the simulation.
- `dx`: Spatial resolution for the simulation.
- `threshold`: Threshold for opinion rigidity.
- `bubble_strength`: Strength of the filter bubble effect.
- `bubble_tolerance`: Tolerance for difference within a filter bubble.

**Equations:**

$$\phi'_{X,i} = \phi_{X,i} + dt \cdot \Big[I_{X,i} + S_{X,i} + B_{X,i} + \\ R_{X,i} + F_{X,i}\Big] \cdot (1 - \text{forgetfulness}[i])$$

The update rule for each opinion state $\phi_{X,i}$ is given by:

$$I_{X,i} = \text{interface energy},$$
$$S_{X,i} = \text{social term},$$
$$B_{X,i} = \text{bias term},$$
$$R_{X,i} = \text{rigidity term},$$
$$F_{X,i} = \text{bubble term}.$$

The simulation runs for a number of steps, updating the opinions according to the above rule and recording the changes in various terms such as bias, social influence, and forgetfulness.

**Gradient Calculation, Evolution Recording**

For each opinion state, the gradient is calculated to determine the rate of change across the spatial dimension. The absolute values of these gradients are recorded to visualize the evolution of opinion changes.

- Time series of changes in bias, social influence, forgetfulness, and opinion rigidity.
- Heatmaps of the gradients of opinion evolution, showing where opinions are changing most rapidly.
- Cumulative distribution functions for different scores, illustrating the distribution of effects like social term and rigidity across the population.

The update function is implemented with the `@jit` decorator from the *Numba* to optimize performance. The simulation iterates over the number of steps, calling the update function and storing the results. It tracks the evolution of opinions and the total changes in various factors.

Diffusion effects(with *Allen − Cahn* term) And The function `update_opinion_dynamics` updates the opinions within a population considering interface energies, diffusion effects.

- $\phi_A, \phi_B, \phi_C$: Arrays representing the opinion states of three different groups.
- `opinion_rigidity`: Represents how strongly individuals resist changing their opinions.

- `bias`: Represents the individual biases towards their current opinion.
- `dt`: The time step for the opinion update.
- `dx`: The spatial step used for calculating diffusion.

**Update Rule**

The update rule for opinion $\phi_X$ at position $i$ is given by the equation:

$$\phi'_{X,i} = \phi_{X,i} + dt \left( \frac{\phi_{X,i-1} - 2\phi_{X,i} + \phi_{X,i+1}}{dx^2} - I_{X,i} + B_X - F_X \cdot \phi_{X,i} \right)$$

where:

- $I_{X,i}$ is the interface energy due to the interaction with other opinions.
- $B_X$ is the bias term for opinion $X$.
- $F_X$ is the forgetfulness factor applied to opinion $X$.

The function returns the updated opinions and the mean diffusion scores for each opinion group.

**Simulation Execution:**

A loop iterates over a number of steps, updating the opinions and recording the mean diffusion scores at each step.

- Time series plots show the evolution of each opinion at each step.
- Heatmaps visualize the evolution of opinions and their gradients over time.
- Cumulative distribution functions ($CDFs$) of mean diffusion scores are plotted to analyze the distribution of diffusion effects across the simulation steps.

**Cumulative Distribution Function ($CDF$)**

The $CDF$ is plotted for mean diffusion scores to understand the probability distribution of these scores over the simulation steps.

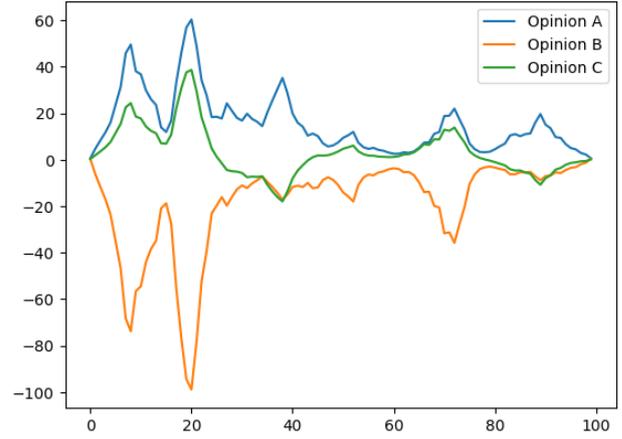

Fig. 2: $\phi_A$, $\phi_B$, $\phi_C$, $Opinion_{Rigidity}$, $Bias$, $t = 1000$

## 3. Discussion

**From Figure 2**

**(1) Characteristics of spatial distribution**

- Opinions A, B, and C show different intensities and peaks at different times.
- Opinion A shows large fluctuations in the initial time step and may be spatially unevenly distributed in the early stages compared to the other two opinions.
- Opinions B and C show more consistent trends and are expected to remain stable over time.

**(2) Trends in Information Propagation and Feedback**

- The large initial fluctuations in Opinion A may indicate that information is propagating rapidly and the feedback loop is active.
- Since all opinions show constant movement between time steps 20 and 60, it can be assumed that information propagation and feedback are relatively stable during this period.

**(3) Trends in internal judgment conditions**

- Judging from the large fluctuations in Opinion A, it is possible that internal judgment conditions are more sensitive than other opinions, or that opinions are adopted or rejected more frequently.

**(4) Parameters controlling the "mixing" of opinions**

- To analyze the "mixing" of opinions, it is necessary to quantify the degree to which individual opinions

overlap, but it is difficult to directly calculate the parameter from this graph alone. This requires additional data to quantify the distance between opinions and the frequency of overlap.

**(5) Trends in phase boundaries between filter bubbles and non-bubble regions**

- Points where opinions cross according to time step (e.g., near time step 40 where opinions B and C cross) indicate that an exchange of opinions may be occurring between filter bubbles and non-bubble regions.

**(6) Tendency of dynamics when different opinions come into contact considering interaction energy**

- The sharp fluctuations at the points of contact between opinions, especially between opinion A and opinion B and opinion C, suggest high interaction energy and strong dynamics between different opinions.
- Peaks and low valleys in opinions could mean that different opinions are influencing each other, and convergence or divergence of opinions may be seen at these points.

## Social Discussion

(1) **Characteristics of spatial distribution**

   Variations in opinion A may reflect a lively debate inspired by a particular social event or media coverage. For example, political developments or scandals may trigger social reactions, and rapidly disseminated information may generate a wave of opinions.

   The consistency exhibited by opinions B and C may reflect more established beliefs or cultural values developed over time. It is suggested that these opinions are based on social consensus rather than new information.

(2) **Information Propagation and Feedback Trends**

   The initial rapid fluctuations shown by Opinion A may point to an information explosion or viral trend on social media. This indicates that emotional responses and immediate sharing are accelerating the dynamics of information propagation.

   The more stable trends exhibited by opinions B and C may indicate that these opinions are generally better understood and represent slowly changing social values and attitudes.

(3) **Trends in internal judgment conditions**

   The fluctuations exhibited by Opinion A may reflect themes that are prone to intensifying debate, e.g., political elections or individuals' opinions on social movements are likely to change.

(4) **"Mixedness" of opinions**

   Areas of overlapping opinions on the graph may indicate a community or forum where people with different opinions and positions interact and influence each other. This may reflect a segment of society where diverse opinions coexist, e.g., multicultural cities or academic settings.

(5) **Information exchange between filter bubble and non-bubble areas.**

   Points of intersection of opinions suggest moments of dialogue between different social bubbles or communities. For example, they may represent the dynamics at a public discussion or a social event where diverse viewpoints are exchanged.

(6) **Interaction energy and dynamics**

   The sharp fluctuations seen between opinions may reflect tensions or clashes between different social groups or opinion clusters. The dynamics may indicate social divisions caused by, for example, election results or policy changes.

**From Figure 3**

(1) **Characteristics of spatial distribution**

   It can be seen that Opinion *A* is more concentrated at certain times of the day (probably after time step 600). This may suggest that opinion *A* has temporarily experienced an event in which it is strongly represented in society. Opinions *B* and *C* maintain a relatively constant distribution over time, while the concentration of *B* has increased over time.

(2) **Tendency of information propagation and feedback**

   The increasing concentration of opinion *A* over time may indicate a tendency for certain information to spread rapidly and for feedback on it to continue. The relatively even spread of opinions *B* and *C* suggests that these opinions have a consistent propagation pattern.

(3) **Trends in internal judgment conditions**

   The concentration of Opinion *A* becomes clearer toward the latter half of the period, which may suggest that an internal judgment mechanism is at work as a response to new information or events.

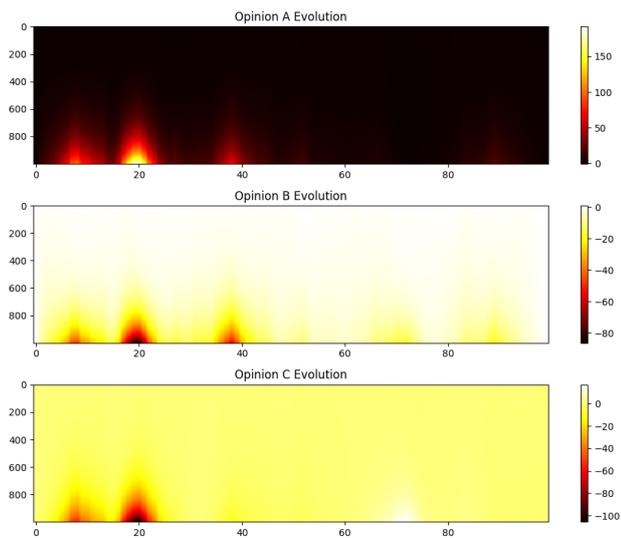

Fig. 3: Opinion $A$, $B$, $C$ Evolution, $t = 1000$

(4) **Parameters controlling the "mixing" of opinions**

While it is difficult to calculate parameters directly from this information, analyzing correlations and temporal continuity among opinions may provide insight into the parameters that regulate the degree of mixing.

(5) **Trends in phase boundaries between filter bubble and non-bubble regions**

The time step of increasing concentration in Opinion $A$ may indicate an increase in filter bubbles. Opinions $B$ and $C$ may indicate less filter bubbles or very extensive bubbles.

(6) **Trends in interaction energy and dynamics**

The abrupt changes seen in Opinion $A$ may indicate strong interactions or clashes between different opinions; $B$ and $C$ may have mild interactions or little interaction; $C$ and $D$ may have very strong interactions or clashes between different opinions; and $A$ and $D$ may have very strong interactions or clashes between different opinions.

(7) **Characteristics as a gradient map**

The heat map of Opinion $A$ has a large gradient of change over time and shows rapid changes during a specific time period. This may indicate that opinions are changing rapidly due to some external factors. Opinions $B$ and $C$ show relatively uniform gradients, indicating that their concentration changes gradually over time.

**Social Discussion**

(1) **The case of Opinion $A$ (e.g., the sudden rise of the environmental protection movement)**

The sudden rise in support for Opinion $A$ over time may represent a phenomenon in which, for example, after a major environmental disaster, opinions about environmental protection rapidly gain support throughout society.

(2) **The case of Opinion $B$ (e.g., a stable but gradually gaining support for a healthy lifestyle)**

Opinion $B$ is gaining support evenly but steadily over time. This may reflect, for example, a situation where interest in healthy lifestyles and organic foods is slowly spreading and gaining support.

(3) **The case of Opinion $C$ (e.g., traditional values that are slowly being lost)**

Opinion $C$ shows a gradual decrease in concentration as the time step progresses. This could suggest, for example, that traditional values and culture are gradually losing favor in modern society.

The trend of the gradient can be considered as follows:

Opinion $A$ tends to increase in slope over time. This may indicate that an event or topic suddenly gains importance in society and rapidly spreads discourse. This type of gradient tends to be seen when social movements or emergencies occur.

Opinions $B$ and $C$ show a more gradual gradient, indicating that social change is gradual. For example, this is how opinions are formed over time, such as economic trends or long-term changes in health awareness.

**From Figure 4**

(1) **Spatial Distribution of Opinion Classes $A$-$C$**

Opinion $A$ shows a sharp gradient at certain time steps, indicating moments of intense change or a strong shift in opinion $A$ at those points in time. Opinion $B$ shows less intense gradients, suggesting more gradual changes in opinion $B$ over time. Opinion $C$ also shows significant gradients at certain time steps, indicating strong shifts similar to opinion $A$ but possibly at different times or with different patterns.

(2) **Trends in Information Propagation and Feedback**

The gradients in opinion $A$ suggest moments of rapid spread or shifts in opinion, possibly due to a feedback loop where a certain trigger caused the opinion to become suddenly more prominent. Opinion *B* and *C*'s gradients suggest more consistent propagation over time, possibly indicating sustained discussions or debates that slowly shift public opinion.

(3) **Trends in Internal Judgment Conditions**

The sharp gradients in opinions *A* and *C* could indicate that the internal conditions or thresholds for change are met abruptly, resulting in sudden shifts in opinion. The smoother gradients in opinion *B* suggest a more continuous reassessment or gradual evolution of internal judgment conditions.

(4) **Parameters Controlling the 'Mix' of Opinions**

These parameters would be related to the rate of change seen in the gradients. A more significant gradient suggests a less mixed opinion, while a smoother gradient suggests a more homogeneous mixing of opinions.

(5) **Interface Trends Between Filter Bubbles and Non-Bubble Regions**

The sharp changes in gradients could indicate the boundaries of filter bubbles where the opinion is either reinforced or rapidly changes due to new information breaking through.

(6) **Dynamics of Interaction Energy When Different Opinions Meet**

The intensity of the gradients at certain points suggests high interaction energy, where a collision of differing opinions might lead to significant shifts or conflicts.

(7) **Characteristics as Gradient Maps**

These gradient maps show where and when the most significant changes in opinion occur. High-intensity areas indicate points of likely social or informational upheaval.

Overall, these gradient maps could be indicative of how public opinion or sentiment about certain topics changes over time. They might reflect real-world events like political campaigns, social movements, or the spread of news stories, where public opinion can shift rapidly in response to new information or due to reinforcing feedback within social or communication networks. The exact dynamics would depend on the context of these opinions and the external factors influencing them.

## Social Discussion

**Impact of Political Events:** During election cycles or important political events, one opinion (e.g., Opinion

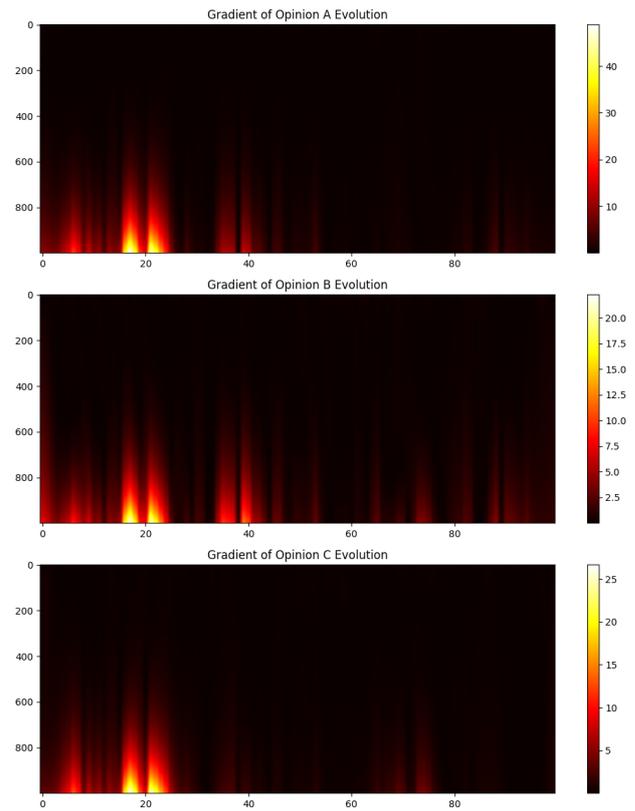

Fig. 4: Gradient of Opinion *A*,*B*, *C* Evolution, *t* = 1000

$\phi_A$) may show a sudden gain or loss of support. This sudden change may reflect the impact of a political speech or scandal.

**Spread of a Social Movement:** A gradual spreading gradient, such as Opinion $\phi_B$, may indicate a social movement gaining power over time. It suggests a tendency to gain acceptance only slowly in the beginning, but eventually spread rapidly.

**Fluctuations in Cultural Trends:** The slope of Opinion $\phi_C$ may indicate a change in a fad or cultural trend. A steeper gradient at a particular time point may indicate increased social interest or rapid adoption of a new trend.

### Hypothetical Considerations

Hypothetical considerations that can be read from the time changes and gradient trends include:

– **The Ever-Changing Flow of Information:** These maps show how rapidly information spreads through social networking sites and news media to influence social opinion. The more rapid the flow of information, the more rapidly social opinion can change.

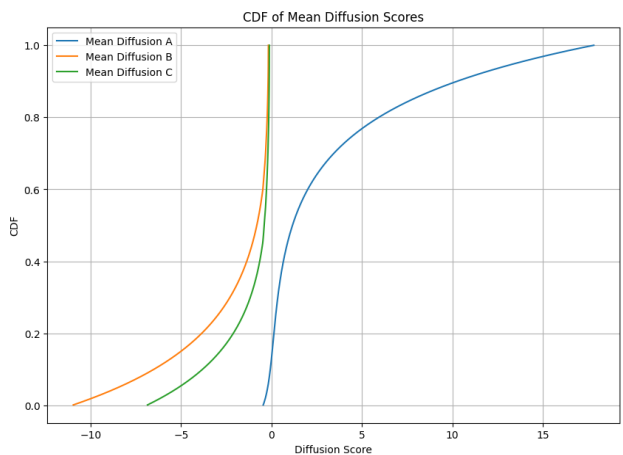

Fig. 5: CDF of Mean Diffusion Scores $A,B$, $Ct = 1000$

- **Impact of Filter Bubbles:** Areas of rapid change in opinion may represent points where filter bubbles burst or, conversely, new bubbles are formed. This may capture the moment when beliefs that are reinforced within a particular community are challenged by outside information.
- **Gradients and Social Conflict:** Steep gradients seen between some opinions indicate areas of particularly active social conflict and debate, and these represent situations where conflicts of opinion are likely to occur.

**From Figure 5**

**(1) Characteristics of Spatial Distribution**

- **Mean Diffusion $\phi_A$:** The distribution extends longer to the right, suggesting that this opinion class may be more widely and uniformly distributed than the other two.
- **Mean Diffusion $\phi_B$:** The CDF rises sharply, suggesting that opinion B is concentrated in a particular area.
- **Mean Diffusion $\phi_C$:** This curve indicates that opinion C is somewhat widely diffused, but not as widely as A.

**(2) Information Propagation and Feedback Tendency**

Information seems to propagate most efficiently within Opinion $\phi_A$. Information propagates within a relatively narrow range for $\phi_B$ and $\phi_C$, and the feedback loop may be strong within those ranges.

**(3) Tendency of internal judgment conditions**

The degree of diffusion of each opinion class may be related to the internal judgment process by which individuals or groups form or change their opinions based on new information; $\phi_A$ may indicate more open judgment conditions, while $\phi_B$ and $\phi_C$ may indicate more strict or limited judgment conditions.

**(4) Parameters for mixing of opinions**

Parameters that control opinion mixing may be inferred from the degree of overlap in diffusion scores between different opinion classes. For example, the area where the $\phi_A$ and $\phi_C$ curves overlap indicates the point where opinion mixing occurs.

**(5) Trend of phase boundary between filter bubble and non-bubble areas**

Areas where the CDF rises sharply indicate that filter bubbles are likely to be present and information is spreading rapidly within those bubbles. Flat areas may indicate slow propagation of information outside the bubbles.

**(6) Dynamics trends when different opinions come into contact**

The dynamics when different opinions come into contact are more pronounced at the point where the curves intersect. This point indicates the point where the different opinion classes are diffused to the same degree and interaction is likely to occur.

While this CDF graph provides important insights into the degree of opinion diffusion and how it proceeds, it should be kept in mind that these interpretations are hypothetical without a specific social context or additional data.

### Social Discussion

**Assumptions of the Speech Case**

- **Mean Diffusion $\mathcal{A}$ (widespread diffusion):** Speech or arguments that are widely accepted in society, such as the growing concern for environmental protection or health, may fit this pattern. Over time, the theme shows a certain resonance with almost all social groups.

- **Mean Diffusion $\mathcal{B}$ (Concentration in a Specific Domain)**: Discourse related to a niche hobby or specific ideology is expected to be rapidly shared within a relatively narrow cluster. For example, information related to a particular subculture or a particular political belief.
- **Mean Diffusion $C$ (limited diffusion)**: More widely diffused but not as widespread as $\mathcal{A}$, this might include new scientific discoveries or limited regional interests.

**Trends as Time-Varying**

- Clusters that grow over time: Over time, a particular opinion or topic may gradually gain social acceptance. For example, a new advance in technology or a cultural trend may gradually gain mainstream acceptance.
- Rapid diffusion: A social or cultural event (e.g., a scandal or urgent news story) can spread rapidly within a particular community, possibly causing a rapid rise, such as Mean Diffusion $\mathcal{B}$.

**Trends in Information Diffusion (Allen-Kahn Term)**

- Information homogenization: Diffusion patterns such as Mean Diffusion $\mathcal{A}$ suggest that information is spread evenly over a wide area and accepted by diverse communities. In this case, the Allen-Kahn term can indicate a tendency for information to be exchanged among different clusters and homogenized over time.
- Localized information enhancement: The Mean Diffusion $\mathcal{B}$ pattern indicates a tendency for information to be locally enhanced within a given community. The Allen-Kahn term may indicate that information exchange within such clusters is very active, but limited between clusters.

## 4. Conclusion

In this paper, we:

(1) Calculate the interaction energies at the phase interfaces:

$$\text{interface\_energy}_{\mathcal{A}}, \text{interface\_energy}_{\mathcal{B}}, \text{interface\_energy}_{C}$$

(2) Social effects:

$$\text{social\_term}_{\mathcal{A}}, \text{social\_term}_{\mathcal{B}}, \text{social\_term}_{C}$$

(3) Application of confirmation bias:

$$\text{bias\_term}_{\mathcal{A}}, \text{bias\_term}_{\mathcal{B}}, \text{bias\_term}_{C}$$

(4) Applying rigidity of opinion:

$$\text{rigidity\_term}_{\mathcal{A}}, \text{rigidity\_term}_{\mathcal{B}}, \text{rigidity\_term}_{C}$$

(5) Taking into account the effect of filter bubbles:

$$\text{bubble\_term}_{\mathcal{A}}, \text{bubble\_term}_{\mathcal{B}}, \text{bubble\_term}_{C}$$

**From Figure 6**

We would like to discuss the results from the CDF of ...

- **Characteristics of Spatial Distribution**
  * *Interface Energy (interaction energy at the phase interface)*: Looking at the CDF for Group 1,

    $$\text{Interface Energy}_{\mathcal{A}}, \text{Interface Energy}_{\mathcal{B}},$$
    $$\text{and Interface Energy}_{C}$$

    are similar, and there is an overall distribution with high score values. This may mean that the energy at which different opinion clusters interact is high and that there is high friction of opinions between clusters.

(1) **Information Propagation and Feedback Tendency**

- *Social Term (Social Influence)*: In the CDF of Group 2, Social Term$_{\mathcal{A}}$ shows a slower rise in CDF than the other two, indicating less overall social influence; Social Term$_{\mathcal{B}}$ rises very rapidly, which may mean strong information feedback in some groups; Social Term$_{C}$ is similar to Social Term$_{\mathcal{B}}$, but and is rising slightly more slowly.

(2) **Trends in Internal Decision Conditions**

- *Bias Term (Confirmation Bias)*: In the Group 3 CDF, Bias Term$_{\mathcal{A}}$ rises most slowly, suggesting the least fixed nature of opinions, while Bias Term$_{\mathcal{B}}$ and Bias Term$_{C}$ rise very steeply, which may indicate a high presence of confirmation bias in the group.

(3) **Parameters Controlling for the "Mixedness" of Opinions**

- *Rigidity Term (Rigidity of Opinions)*: The CDFs for Group 4 are related to the Rigidity Term: the CDF for Rigidity Term$_{\mathcal{A}}$ is the flattest, indicating that opinions are flexible and tend to mix; Rigidity Term$_\mathcal{B}$ and Rigidity Term$_C$ are very steep, meaning that opinions are more rigid and less likely to mix. To calculate the parameters that control mixing, we need to quantify the range of score values where information is actively exchanged based on the shape of these CDFs.

(4) **Information Exchange Between Filter Bubble and Non-Bubble Regions**
   – *Bubble Term (Effect of Filter Bubble)*: In the Group 5 CDF, Bubble Term$_\mathcal{A}$ and Bubble Term$_C$ almost overlap, indicating that information bubbles are similar within each cluster; Bubble Term$_\mathcal{B}$ shows a slightly different distribution than these, which may mean that there are differences in information exchange between bubble and non-bubble regions.

(5) **Interaction Energy and Opinion Dynamics**
   – Looking at the CDF of interaction energy, the dynamics when different opinions come into contact shows that the energy between clusters of opinions is high, as seen in Group 1, suggesting a situation where conflicts and arguments are likely to occur. High energy is less likely to result in a change of opinion and may intensify conflict.

## Social Discussion

### Interaction energy at the phase interface

This energy suggests how active the exchange of opinions and beliefs is between different social clusters, or how prone to conflict they are. For example, the phase interface energy is likely to be high between groups that are clearly divided, such as political left-right or religious conflicts. This means that when their opinions come into contact with each other, either the debate becomes more active or the conflict intensifies.

### Rigidity of Opinions

Rigidity of opinion indicates that opinions within a group are unlikely to change. In the real world, this is often seen in groups with strong ideologies and communities with a strong sense of in-group solidarity. For example, an extreme political group or a strict religious community may fall into this category.

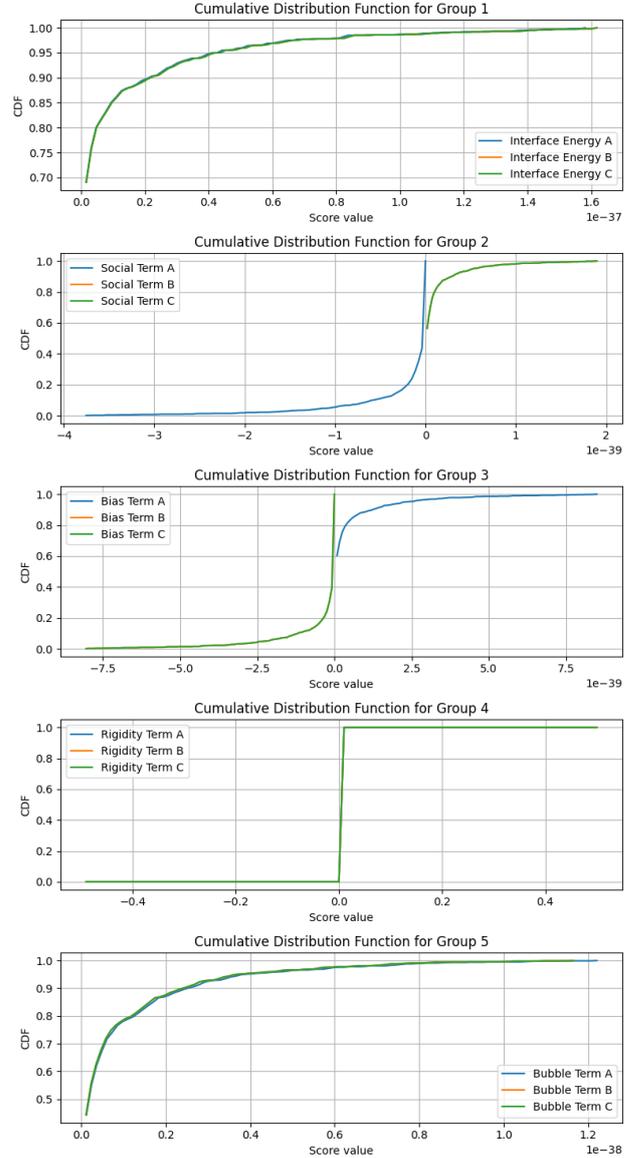

Fig. 6: Cumulative Distribution Function for Group $CDF$ Score value $A, B, Ct = 1000$

**Confirmation Bias**

Confirmation bias describes the tendency of individuals to accept only information that supports their beliefs and ignore or deny conflicting information. Social media echo chambers, in which only voices of agreement reverberate, are a classic example of this phenomenon. Social divisions can deepen as people are exposed only to sources that reinforce their views.These factors play an important role in the dynamics between groups and in the formation of social discourse. For example, political debates during elections, social movements and protests, religious dialogues, and scientific discussions (e.g., debates about vaccines or climate change) are all possible. Understanding how the above parameters are affected in these debates would be helpful in developing strategies to promote dialogue and reduce fragmentation.

# Aknowlegement

The author is grateful for discussion with Prof. Serge Galam.This research is supported by Grant-in-Aid for Scientific Research Project FY2 019-2021, Research Project/Area No. 19K04881, "Construction of a new theory of opinion dynamics that can describe the real picture of society by introducing trust and distrust".